# Extended Depth-range Dual-wavelength Interferometry Based on Iterative Two-step Temporal Phase-unwrapping

Minmin Wang[a], Canlin Zhou[a]*, Shuchun Si[a], XiaoLei Li[b], Zhenkun Lei[c]**, YanJie Li[d]

[a]*School of Physics, Shandong University, Jinan, China;* [b] *School of Mechanical Engineering, Hebei University of Technology, Tianjin, China;* [c]*Department of engineering mechanics, Dalian University of Technology, Dalian, China;* [d]*School of civil engineering and architecture, Jinan University, Jinan, China*

*Corresponding author: Tel: +8613256153609; E-mail address: canlinzhou@sdu.edu.cn

**Corresponding author: Tel: +8615841175236; E-mail address: leizk@163.com

# Extended Depth-range Dual-wavelength Interferometry Based on Iterative Two-step Temporal Phase-unwrapping

Phase retrieval is one of the most challenging processes in many interferometry techniques. To promote the phase retrieval, Xu et. al [X. Xu, Y. Wang, Y. Xu, W. Jin. 2016] proposed a method based on dual-wavelength interferometry. However, the phase-difference brings large noise due to its low sensitivity and signal-to-noise ratio (SNR). Beside, special phase shifts are required in Xu's method. In the light of these problems, an extended depth-range dual-wavelength phase-shifting interferometry is proposed. Firstly, the least squares algorithm is utilized to retrieve the single-wavelength phase from a sequence of N-frame simultaneous phase-shifting dual-wavelength interferograms (SPSDWI) with random phase shifts. Then the phase-difference and phase-sum are calculated from the wrapped phases of single wavelength, and the iterative two-step temporal phase-unwrapping is introduced to unwrap the phase-sum, which can extend the depth-range and improve the sensitivity. Finally, the height of objects is achieved. Simulated experiments are conducted to demonstrate the superb precision and overall performance of the proposed method.

## 1. Introduction

Interferometry has been widely employed in optical phase and surface topography measurement due to its advantages of high resolution, non-contact, and noise resistance (1-3). Over the years, many phase demodulation methods have been proposed to retrieve the phase from interferograms. Among these methods, the dual-wavelength interferometry (DWI) not only keeps the merits of traditional single-wavelength interferometry, but also can produce a non-wrapped phase-difference by the simple subtraction between two wrapped single-wavelength phases (4). Hence the DWI have been naturally investigated by researchers and there are several methods of DWI developed for phase retrieval, most of which can be classified into two categories: spatial-Fourier-transform (SFT) based phase retrieval (5-8) and temporal-phase-shifting (TPS) based phase retrieval (9-11).

The SFT method is a better choice for scenes that require real-time and high speed as the wrapped phase can be retrieved from only a single-frame dual-wavelength interferogram through Fourier transform. Onodera et. al (7) investigated a two-wavelength interferometry based on a Fourier-transform method and analyzed the phase error caused by the difference between modulation intensities at two wavelengths. Min et. al (8) proposed a dual-wavelength digital holographic microscopy with a slightly off-axis configuration, where the high wavelength selectivity of the Bayer mosaic filtered color CCD camera is used. Kühn et. al (12) developed a technique to perform two-wavelengths digital holographic microscopy measurements with a single hologram acquisition. The real-time dual-wavelength imaging is realized by using two reference waves with different wavelengths and propagation directions for the hologram recording. Ichioka et. al (13) proposed a digital holographic configuration utilized for single-shot, dual-wavelength, off-axis geometry and imaging polarimetry. The common problem of these SFT methods is that their measuring accuracy is greatly influenced by the level of random noise, the filtering window, and the Gibbs boundary effects.

The TPS methods reveal higher accuracy to retrieve the phase compared with the SFT method. Double-wavelength interferometry was improved by Polhemus [14] and later on by Cheng [15] using digital phase-shifting phase-demodulation Afterwards Kakue et. Al (16) applied the image-reconstruction algorithm of parallel two-step phase-shifting digital holography to the hologram so as to propose an algorithm that can improve the quality of the reconstructed image from the single hologram. Abdelsalam and Kim (10) described a configuration used for two-wavelength phase-shifting in-line interferometry based on polarizing separation. This method needs a sequence of phase-shifting interferograms for each single-wavelength, which is rather complicated and may consume a significant amount of time. Huang et. al (17) developed a dual-wavelength interferometry based on the spatial carrier-frequency phase-shifting method, where the least squares method is iteratively utilized. Thus it also has a heavy computational time and memory requirements. Zhang et. al (18) proposed a phase-shifting dual-wavelength interferometry based on the two-step demodulation algorithms. The phase is retrieved through five frames of simultaneous phase-shifting dual-wavelength interferograms with the special phase shifts by using subtraction and the Gram-Schmidt two-step demodulation algorithm. Xu et. al (19) improved Zhang's approach, and proposed an approach of quantitative phase extraction based on two intensities with dual wavelength after filtering the corresponding dc terms for each wavelength. However, a special phase shift is required in Xu's method. Besides, the previous DWI including Xu's method rely only on the phase-difference. Since the phase-difference has low sensitivity and signal-to-noise ratio (SNR), the retrieved phase may be quite noisy.

In this paper, we present an extended depth-range dual-wavelength interferometry for phase retrieval based on combined the iterative two-step temporal phase unwrapping and the least squares method. It is an improvement of Xu's method (19). The limitation of being exclusively applicable to interferograms with special phase shifts is solved by using the least squares method. By introducing in the phase-sum and the iterative two-step temporal phase unwrapping, the sensitivity and SNR is greatly improved. The suitability and superiority of the proposed method is illustrated on simulated experiments.

## 2. Theory

Assume that two lasers with different wavelengths travel the same inline phase-shifting interference system at the same time. The phase shifts of two reference waves come into being simultaneously by a piezoelectric transducer (PZT). A series of interferograms are then captured by a monochrome CCD, which are mathematically described as

$$I_n(x,y) = a(x,y) + \sum_{i=1}^{2} b_i(x,y)\cos[\alpha_i x + \varphi_i(x,y) + \delta_{i,n}] \qquad (1)$$

where $a(x, y)$ is the background intensity; the subscript $i = 1, 2$ denotes two single wavelengths; $b_i(x,y)$ is the intensity modulation amplitudes at a wavelength of $\lambda_i$; $\delta_{i,n}$ is the phase step at $\lambda_i$. $\varphi_i(x,y) = 2\pi h(x,y)/\lambda_i$ is the phase to be retrieved at $\lambda_i$; $h(x, y)$ is optical path difference. $n = 1, 2, 3...N$, $n$ is the image index, and $N$ is the total number of images. The relationship between the phase shifts and the wavelengths is given by

$$\delta_{1,n} / \delta_{2,n} = \lambda_2 / \lambda_1 \qquad (2)$$

*2.1 Xu's method*

The wrapping-free phase extraction method proposed by Xu et. al has been thoroughly described and discussed (19). This section only introduces its main idea briefly. In Xu's method, firstly, the corresponding dc terms are eliminated by solving the intensity difference between $I_1$ and other intensity values. Then, the phase shifts of $\delta_{1,n}$ ($\delta_{2,n}$) are set as $2\pi$ and $4\pi$, thus two intensities of $\lambda_2$ ($\lambda_1$) can be separated, and the wrapped phase of $\varphi_2$ ($\varphi_1$) can be retrieved. After determining the single wavelength phase, the wrapped phase-difference is obtained by doing subtraction between two single wavelength phase, which is then unwrapped to further calculate the height of the object. Clearly, Xu's method is only applicable for interferograms with special phase shifts, which means it will fail once the phase-shifting inducer (i.e. PZT) is not capable of producing a specific phase shift. Besides, the sensitivity and SNR of the phase-difference is rather low, so Xu's method is easily destroyed by the phase errors (harmonics and noise).

*2.2 proposed method*

To solve these problems, we present a robust dual-wavelength simultaneous phase-shifting interferometry that is based on combined the least squares algorithm and the iterative two-step temporal phase-unwrapping method.

**2.2.1 the generalized phase shifting algorithm**

To overcome the limitation of phase shifts without sacrificing the accuracy, here, we review the generalized phase shifting algorithm (20-23). Eq. (1) can be

rewritten as

$$I_n(x,y) = a(x,y) + b_1(x,y)\cos[\varphi_1(x,y)]\cos(\delta_{1,n}) - b_1(x,y)\sin[\varphi_1(x,y)]\sin(\delta_{1,n}) \\ + b_2(x,y)\cos[\varphi_2(x,y)]\cos(\delta_{2,n}) - b_2(x,y)\sin[\varphi_2(x,y)]\sin(\delta_{2,n})$$ (3)

Instead of making the subtraction operation between the intensity distribution of two interferograms, we define a new set of variables as $b(x,y)=b_1(x,y)\cos[\varphi_1(x,y)]$, $c(x,y)=-b_1(x,y)\sin[\varphi_1(x,y)]$, $d(x,y)=b_2(x,y)\cos[\varphi_2(x,y)]$, and $e(x,y)=-b_2(x,y)\sin[\varphi_2(x,y)]$, Eq. (3) can be described as

$$I_n(x,y) = a(x,y) + b(x,y)\cos(\delta_{1,n}) + c(x,y)\sin(\delta_{1,n}) \\ + d(x,y)\cos(\delta_{2,n}) + e(x,y)\sin(\delta_{2,n})$$ (4)

The deviation square sum for all $N$ fringe patterns can be expressed as

$$E(x,y) = \sum_{n=1}^{N} \left[ \begin{array}{c} a(x,y) + b(x,y)\cos(\delta_{1,n}) + c(x,y)\sin(\delta_{1,n}) \\ + d(x,y)\cos(\delta_{2,n}) + e(x,y)\sin(\delta_{2,n}) - I_n'(x,y) \end{array} \right]^2$$ (5)

where $I_n'$ is the intensity of interferograms measured in the experiments. According to the principle of the least squares method, the following extreme value condition should be satisfied

$$\frac{\partial E(x,y)}{\partial a(x,y)}=0, \quad \frac{\partial E(x,y)}{\partial b(x,y)}=0, \quad \frac{\partial E(x,y)}{\partial c(x,y)}=0, \quad \frac{\partial E(x,y)}{\partial d(x,y)}=0, \quad \frac{\partial E(x,y)}{\partial e(x,y)}=0$$ (6)

In this way, five unknowns of $a(x, y)$, $b(x, y)$, $c(x, y)$, $d(x, y)$, $e(x, y)$ can be resolved simultaneously from these equations. Solve Eq. (6) to get the following matrices

$$X = U^{-1}Q$$ (7)

where

$$U = \begin{bmatrix} N & \sum_{n=1}^{N} c_{1,n} & \sum_{n=1}^{N} s_{1,n} & \sum_{n=1}^{N} c_{2,n} & \sum_{n=1}^{N} s_{2,n} \\ \sum_{n=1}^{N} c_{1,n} & \sum_{n=1}^{N} c_{1,n}^2 & \sum_{n=1}^{N} c_{1,n}s_{1,n} & \sum_{n=1}^{N} c_{1,n}c_{2,n} & \sum_{n=1}^{N} c_{1,n}s_{2,n} \\ \sum_{n=1}^{N} s_{1,n} & \sum_{n=1}^{N} c_{1,n}s_{1,n} & \sum_{n=1}^{N} s_{1,n}^2 & \sum_{n=1}^{N} c_{2,n}s_{1,n} & \sum_{n=1}^{N} s_{1,n}s_{2,n} \\ \sum_{n=1}^{N} c_{2,n} & \sum_{n=1}^{N} c_{1,n}c_{2,n} & \sum_{n=1}^{N} c_{2,n}s_{1,n} & \sum_{n=1}^{N} c_{2,n}^2 & \sum_{n=1}^{N} c_{2,n}s_{2,n} \\ \sum_{n=1}^{N} s_{2,n} & \sum_{n=1}^{N} c_{1,n}s_{2,n} & \sum_{n=1}^{N} s_{1,n}s_{2,n} & \sum_{n=1}^{N} c_{2,n}s_{2,n} & \sum_{n=1}^{N} s_{2,n}^2 \end{bmatrix}$$

$$X = [a(x,y) \ b(x,y) \ c(x,y) \ d(x,y) \ e(x,y)]^T$$

$$Q = [\sum_{n=1}^{N} I_n \quad \sum_{n=1}^{N} I_n c_{1,n} \quad \sum_{n=1}^{N} I_n s_{1,n} \quad \sum_{n=1}^{N} I_n c_{2,n} \quad \sum_{n=1}^{N} I_n s_{2,n}]^T \tag{8}$$

where $c_{i,n} = \cos\delta_{i,n}$ and $s_{i,n} = \sin\delta_{i,n}$ (i = 1, 2).

By taking $N = 5$ and solving Eq. (7) through five interferograms with arbitrary phase shifts, the wrapped phases can be obtained by

$$\varphi_1^w(x,y) = \arctan(-\frac{c(x,y)}{b(x,y)}), \quad \varphi_2^w(x,y) = \arctan(-\frac{e(x,y)}{d(x,y)}) \tag{9}$$

**2.2.2 Phase-difference and phase-sum estimations**

The phase-difference and highly wrapped phase-sum can be calculated by

$$\varphi_d^w(x,y) = \varphi_1^w(x,y) - \varphi_2^w(x,y) \tag{10a}$$

$$\varphi_s^w(x,y) = \varphi_1^w(x,y) + \varphi_2^w(x,y) \tag{10b}$$

If the wavelength of $\lambda_1$ and $\lambda_2$ are chosen close, their phase-difference $\varphi_d$ is continuous after using simple addition (given in Eq. (11)) since the synthetic wavelength is much larger (20).

$$\varphi_d(x,y) = \begin{cases} \varphi_d^w(x,y) & \text{while } \varphi_d^w(x,y) \geq 0 \\ \varphi_d^w(x,y) + 2\pi & \text{else} \end{cases} \tag{11}$$

Plenty of dual-wavelength interferometry including Xu's method rely only on the phase-difference as the non-wrapped phase-difference contains the searched phase-profile and is already unwrapped. However, we think that the phase-sum should also be included to improve the sensitivity and precision of the retrieved phase (25). The advantages of the phase-sum are illustrated as follows.

First, the sensitivity of the phase-sum and phase-difference is compared. The equivalent fringe periods of the phase-difference and phase-sum are expressed as

$$\lambda_s = \frac{\lambda_1 \lambda_2}{\lambda_1 + \lambda_2}; \quad for\ \varphi_1 + \varphi_2 \tag{12a}$$

$$\lambda_s = \frac{\lambda_1 \lambda_2}{|\lambda_1 - \lambda_2|}; \quad for\ \varphi_1 - \varphi_2 \tag{12b}$$

If the wavelength of $\lambda_1$ and $\lambda_2$ are chosen close enough, namely $\lambda_1 \approx \lambda_2$, $\lambda_d$ is much larger than $\lambda_1$ or $\lambda_2$, while $\lambda_s$ is much shorter than either $\lambda_1$ or $\lambda_2$. For that reason, the phase-difference has low sensitivity while the phase-sum has high sensitivity. Then the sensitivity gain $G$ between the phase-difference and phase-sum is

$$G = \frac{\lambda_1 + \lambda_2}{|\lambda_1 - \lambda_2|} = \frac{\lambda_s}{\lambda_d} \tag{13}$$

Next, we illustrate the superiority of the phase-sum by analyzing and contrasting the SNR of the phase-sum and phase-difference. In practice white Gaussian noise may pollute the demodulated phases $\varphi_1$ and $\varphi_2$. Thus a measuring phase-noise should be added to the demodulated phases (26). This means the demodulated phases can be expressed as

$$\varphi_1(x, y) = 2\pi h(x, y) / \lambda_1 + e_1(x, y) \tag{14a}$$

$$\varphi_2(x, y) = 2\pi h(x, y) / \lambda_2 + e_2(x, y) \tag{14b}$$

where $e_1(x, y)$ and $e_2(x, y)$ are two noise samples. The phase-difference and phase-sum

are re-expressed as

$$\varphi_d(x,y) = 2\pi h(x,y)(\lambda_2 - \lambda_1)/\lambda_1\lambda_2 + e_2(x,y) - e_1(x,y)$$

$$\varphi_s(x,y) = 2\pi h(x,y)(\lambda_2 + \lambda_1)/\lambda_1\lambda_2 + e_2(x,y) + e_1(x,y) \tag{15}$$

Then the SNR for $\varphi_d$ and $\varphi_s$ are

$$\tau_d = \frac{4\pi^2(\lambda_2-\lambda_1)^2/\lambda_1^2\lambda_2^2 \iint\limits_{(x,y)\in\Omega} |h(x,y)|^2 d\Omega}{\iint\limits_{(x,y)\in\Omega} |e_2(x,y)-e_1(x,y)|^2 d\Omega} \tag{16a}$$

$$\tau_s = \frac{4\pi^2(\lambda_2+\lambda_1)^2/\lambda_1^2\lambda_2^2 \iint\limits_{(x,y)\in\Omega} |h(x,y)|^2 d\Omega}{\iint\limits_{(x,y)\in\Omega} |e_2(x,y)+e_1(x,y)|^2 d\Omega} \tag{16b}$$

where $\Omega$ is the well-defined fringe-data. As both noise samples are generated by the same Gaussian zero-mean stationary random process, in the average, the energy of $e_2(x,y)+e_1(x,y)$ and $e_2(x,y)-e_1(x,y)$ are equal (26). Therefore, the SNR gain between $\varphi_s$ and $\varphi_d$ is,

$$\frac{\tau_s}{\tau_d} = \left(\frac{\lambda_2+\lambda_1}{\lambda_2-\lambda_1}\right)^2 = G^2 \tag{17}$$

The phase-sum has $G^2$ higher SNR than the phase-difference. Given the theoretical analysis mentioned above, the phase-sum should be introduced to retrieve the phase with high sensitivity and high precision.

**2.2.3 the 2-step temporal phase unwrapping algorithm**

The two-step temporal phase unwrapping algorithm is used to retrieve the continuous phase in this paper. The non-wrapped phase-difference $\varphi_d$ is utilized as the first estimation to temporarily unwrap the phase-sum $\varphi_s$ (27-28), then we can obtain the unwrapped phase as

$$\varphi_s(x,y) = G\varphi_d(x,y) + W\left[\varphi_s^w(x,y) - G\varphi_d(x,y)\right] \tag{18}$$

where $W$ is the wrapping phase operator, and $\varphi_s^w(x,y) = W[\varphi_s(x,y)]$. Eq. (18) is effective, only when the following condition is fulfilled,

$$[\varphi_s(x,y) - G\varphi_d(x,y)] \in (-\pi, \pi) \tag{19}$$

**2.2.3 the iterative two-step temporal phase unwrapping algorithm**

We know from Eq. (17) that the greater the sensitivity gain $G$ is, the greater the SNR gain of the demodulation phase. Therefore, in order to improve the measurement accuracy, the $G$ value should be chosen as large as possible. However, as can be seen from Eq. (18), the phase $\varphi_d$ is scaled up by $G$ to further unwrap the phase $\varphi_s$. Thus the noise in the phase $\varphi_d$ is also magnified. Besides, when $\lambda_1$ and $\lambda_2$ are chosen close to each other, $G$ is very large, so it may be hard to satisfy the condition shown in Eq. (19), so noise may destroy the phase retrieval process. To increase the sensitivity gain $G$ as well as reducing the noise, here, we utilized the iterative two-step temporal phase unwrapping algorithm. First, the phase $\varphi_2^w$ is unwrapped by the phase-difference $\varphi_d$ using Eq. (20a).

$$\varphi_2(x,y) = G_1\varphi_d(x,y) + W\left[\varphi_2^w(x,y) - G_1\varphi_d(x,y)\right] \tag{20a}$$

$$\varphi_s(x,y) = G_2\varphi_2(x,y) + W\left[\varphi_s^w(x,y) - G_2\varphi_2(x,y)\right] \tag{20b}$$

The non-wrap phase $\varphi_2$ is then utilized to unwrap the phase-sum $\varphi_s^w$ using Eq. (20b). The sensitivity gain for this case is $G_1 = \lambda_2/\lambda_d$ and $G_2 = \lambda_s/\lambda_2$. In this way, the continuous phase-sum is obtained with high sensitivity gain $G = G_1 \times G_2$ and high SNR.

Finally, the thickness of the phase object can be obtained by Eq. (21).

$$h(x,y) = \varphi_s(x,y)\lambda_s / 2\pi \tag{21}$$

Obviously, using the proposed method, the special phase shifts of the fringe patterns are no longer needed, and the sensitivity and SNR are improved.

The main steps of the proposed algorithm can be summarized as follows:

(1) capture a sequence of N-frame ( $N \geq 5$ ) SPSDWI with random phase shifts;

(2) calculate the wrapped phase-maps $\varphi_1^w$ and $\varphi_2^w$ of the single wavelength from fringe patterns using the generalized phase-shifting algorithm;

(3) compute the phase-difference $\varphi_d$ and the highly-wrapped phase-sum $\varphi_s^w$ from the wrapped phase-maps $\varphi_1^w$ and $\varphi_2^w$ according to Eq. (10) and Eq. (11);

(4) unwrap $\varphi_2^w$ with $\varphi_d$ by Eq. (20a), and obtain the continuous phase $\varphi_2$;

(5) unwrap $\varphi_s^w$ with $\varphi_2$ by Eq. (20b), and obtain the continuous phase data $\varphi_s$ with a high sensitivity gain $G$ and high SNR;

(6) obtain the thickness of the phase object from the continuous phase $\varphi_s$ by Eq. (21).

These are the main procedures of the proposed algorithm with larger sensitivity gain $G$ by cascading with 2 level different sensitivities ($G = G_1 \times G_2$). Some following simulated experiments are used to verify the proposed algorithm.

## 3. Simulated experiments and result analysis

To demonstrate the feasibility of the proposed method, the simulated experiments are carried out in MATLAB. Five-frame interferograms with size of 256 × 256 pixels are generated, in which the background is $a(x,y) = 120 \times e^{-0.05 \times (x^2+y^2)}$ and the modulation amplitudes are $b_1(x,y) = 50 \times e^{-0.04 \times (x^2+y^2)}$ and $b_2(x,y) = 60 \times e^{-0.05 \times (x^2+y^2)}$ at 532 nm and 632.8 nm, respectively. The synthetic beat wavelength of the phase-difference is equal to 3339.78 nm, while the equivalent wavelength of the phase-sum is equal to 289.02 nm. A spherical cap with the height of $h_{set} = 480 \times 10^{-6} \times (x^2 + y^2)$, in which -1.27 mm ≤ x, y ≤ 1.28 mm is employed as the measured object. White noise with a variance of 0.6 and a mean of 0 was added to all the five fringe patterns to test the robustness of the proposed method. The phase shifts for the fringe patterns are chosen as $\delta_{\lambda1,1}$=0, $\delta_{\lambda1,2}$=2π, $\delta_{\lambda1,3}$=4π, $\delta_{\lambda1,4}$=2π, $\delta_{\lambda1,5}$=4π at $\delta_{\lambda1,n}$, and $\delta_{\lambda1,n}\lambda_1/\lambda_2$ at $\delta_{\lambda2,n}$, so they are available for both Xu's method and the proposed method, as shown in Fig. 1.

First, we utilized Xu's method to process the simulated patterns. The wrapped single wavelength phases can be calculated from these phase-shifting interferograms using Eq. (9), as shown in Figures 2(a) and 2(b). Figure 3(a) shows the phase-difference between those two wrapped single wavelength phases. As there is no explanation about how to retrieve the continuous phase-difference from the wrapped phase-difference in Xu's method (19), we utilized the *Unwrap* function in MATLAB, and the non-wrapped phase-difference is shown in Figure 3(b). Finally, the thickness of the phase object was determined, as shown in Figure 3(c). It is easy to see that Xu's method accommodates noise with the RMSE of 0.0053 um. Since the sensitivity of the phase-difference is far too small , it is almost impossible for this case to accurately calculate the thickness distribution of the phase object.

Next, we employed the proposed method. The least squares method is utilized to estimate the wrapped single wavelength phases. Then except for the phase-difference, the highly-wrapped phase-sum is obtained, as shown in Figure 4(a). Using the continuous phase-difference to unwrap the phase-sum by the 2-step temporal phase unwrapping algorithm (Eq. 18), the retrieved phase and its thickness are shown in Figure 4(b) and 4(c), respectively. As the sensitivity gain $G = 3339.78/289.02 = 11.56$ is relatively large in this case, the condition shown in Eq. (19) may not be satisfied and the noise in the non-wrapped phase-difference can also be amplified, so many errors are apparent on the results with the RMSE of 0.0188 um. Therefore we introduced the iterative 2-step temporal phase unwrapping algorithm. The non-wrapped phase-difference is firstly employed to unwrap the single wavelength phase at 632.8 nm with the sensitivity gain $G_1 = 3339.78 / 632.8 = 5.28$, as shown in Figure 5(a). The phase-sum is further unwrapped using the continuous single wavelength phase with the sensitivity gain $G_2 = 632.8 / 289.02 = 2.19$, as shown in Figure 5(b), of which the thickness is shown in Figure 5(c). In this way, the sensitivity gain is cascaded with 2 level different sensitivities $G_1 \times G_2$. Thus, the noise is reduced as $G_1$ is relatively small. By applying the proposed method to the wrapped phase maps, it acts to significantly reduce the error with the RMSE of $9.21 \times 10^{-6}$ um.

To better illustrate the difference, Figures 6 show the horizontal cross sections of the retrieved surfaces along the middle row. Through comparing Figures 6, we can see that the proposed method is far more robust to noises when utilizing the iterative two-step temporal phase unwrapping as we have 11.56-times ($G = G_1 \times G_2 = 5.28 \times 2.19 = 11.56$) more sensitivity in the phase-sum compared to the phase-difference.

## 4. Conclusion

In this paper, we presented an extended depth-range dual-wavelength interferometry for phase retrieval based on combined the iterative two-step temporal phase unwrapping and the least squares method. It is an extension of Xu's method. Comparing with Xu's method, the proposed method solves the limitation of being exclusively applicable to interferograms with special phase shifts by utilizing the least squares method and generalized phase shifting algorithm. Besides, by introducing in the phase-sum and the iterative two-step temporal phase-unwrapping, the proposed method greatly extends the depth-range and improves the sensitivity, meanwhile reducing the noise impact. Based on simulated experiments presented in Section 3, the high precision of the proposed method have been successfully demonstrated and confirmed.

**Acknowledgment**

This work was supported by the [National Natural Science Foundation of China] under Grant [number 11672162,11302082 and 11472070]. The support is gratefully acknowledged.

**Figure captions**

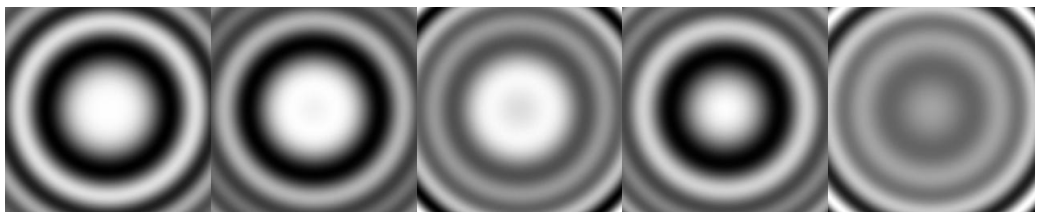

Figure 1. Simulated interferograms with different phase shifts:

(a) 0 at 532 and 632.8 nm;

(b) $2\pi$ at 532 nm;

(c) $4\pi$ at 532 nm;

(d) $2\pi$ at 532 nm;

(e) $4\pi$ at 532 nm.

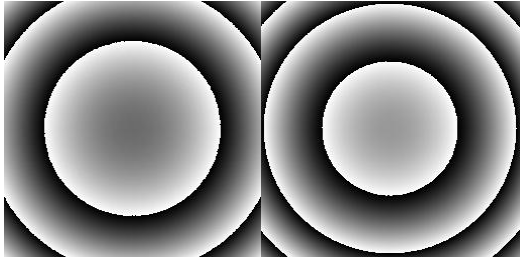

Figure 2. Wrapped phases:

(a) at 532 nm;

(b) at 632.8 nm.

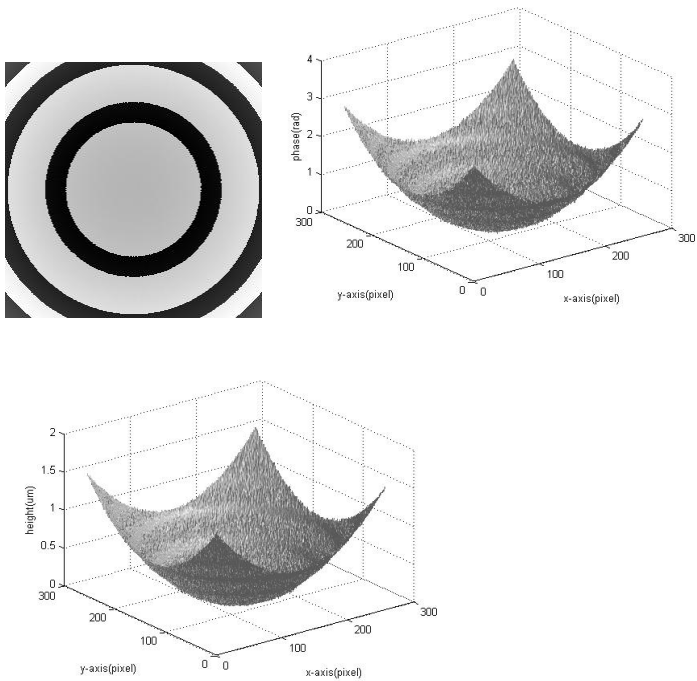

Figure 3. Results obtained by Xu's method:

(a) the wrapped phase-difference;

(b) the retrieved non-wrapped phase-difference using Xu's method;

(c) the thickness of the phase object shown in Figure 3(b).

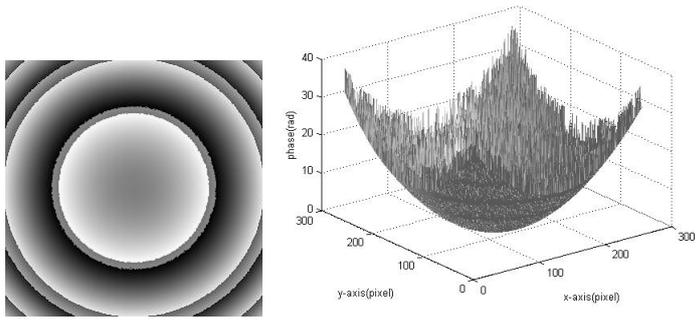

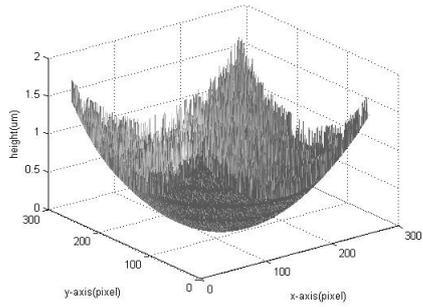

Figure 4. Results obtained by the two-step temporal phase-unwrapping:

(a) the highly-wrapped phase-sum;

(b) the retrieved non-wrapped phase sum using Xu's method;

(c) the thickness of the phase object shown in Figure 4(b).

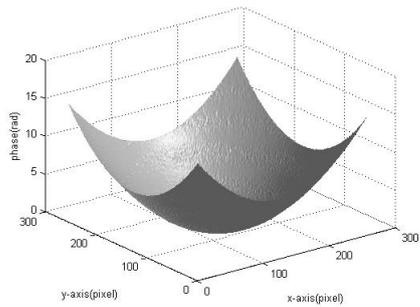 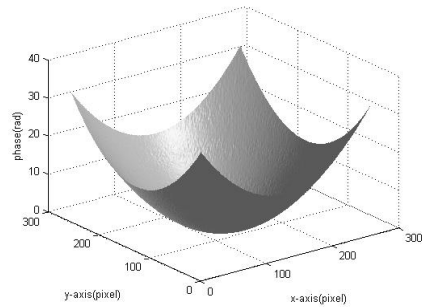

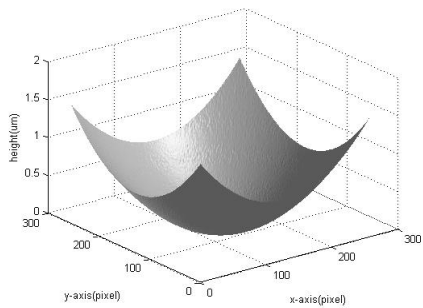

Figure 5. Results obtained by the iterative two-step temporal phase-unwrapping:

(a) the retrieved non-wrapped single wavelength phase at 632.8 nm;

(b) the retrieved non-wrapped phase sum using the iterative two-step temporal phase-unwrapping;

(c) the thickness of the phase object shown in Figure 5(b).

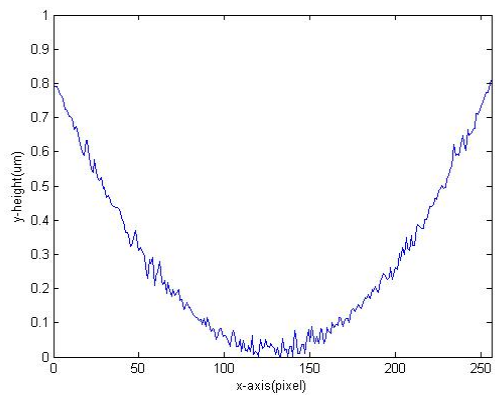

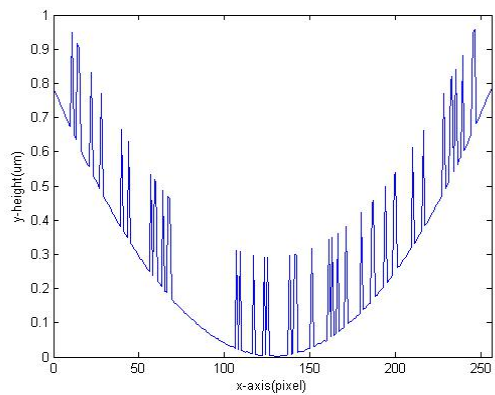

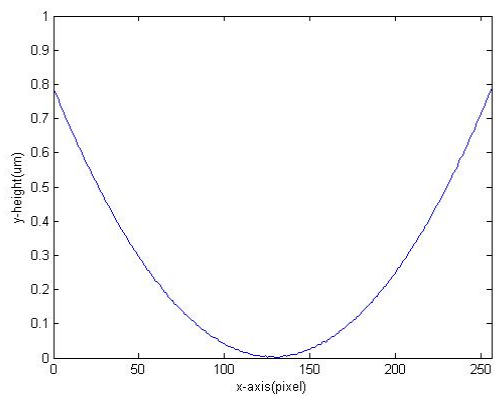

Figure 6. The horizontal cross sections of the retrieved surfaces using:

(a) Xu's method;

(b) the two-step temporal phase-unwrapping;

(c) the iterative two-step temporal phase-unwrapping.